# A unified approach for morphometrics and functional data analysis with machine learning for craniodental shape quantification in shrew species

Aneesha Balachandran Pillay[1], Dharini Pathmanathan[2], Sophie Dabo-Niang[3], Arpah Abu[4], and Hasmahzaiti Omar[5]

[1,2]Institute of Mathematical Sciences, Faculty of Science, Universiti Malaya, Kuala Lumpur, Wilayah Persekutuan Kuala Lumpur, Malaysia.

[3]Laboratoire Paul Painlevé UMR CNRS 8524, INRIA-MODAL, Université de Lelle, France.

[4,5]Institute of Biological Sciences, Faculty of Science, Universiti Malaya, Kuala Lumpur, Wilayah Persekutuan Kuala Lumpur, Malaysia.

17043604@siswa.um.edu.my[1], sophie.dabo@univ-lille.fr[3], arpah@um.edu.my[4], zaiti_1978@um.edu.my[5]

Corresponding author: dharini@um.edu.my[2]

## ABSTRACT

*This work proposes a functional data analysis approach for morphometrics with applications in classifying three shrew species (S. murinus, C. monticola and C. malayana) based on the images. The discrete landmark data of craniodental views (dorsal, jaw and lateral) are converted into continuous curves where the curves are represented as linear combinations of basis functions. A comparative study based on four machine learning algorithms such as naïve Bayes, support vector machine, random forest, and generalized linear models was conducted on the predicted principal component scores obtained from the FDA approach and classical approach (combination of all three craniodental views and individual views). The FDA approach produced better results in separating the three clusters of shrew species compared to the classical method and the dorsal view gave the best representation in classifying the three shrew species. Overall, based on the FDA approach, GLM of the predicted PCA scores was the most accurate (95.4% accuracy) among the four classification models.*

## INTRODUCTION

Morphometric approaches have been widely used in the study of organisms to facilitate the analysis of quantitative variation in form (Roth, 2000). Geometric morphometrics (GM) is a powerful tool in quantifying and identifying biological shapes as well as variations in living organisms. This method supersedes traditional morphometrics (TM) which uses linear measurements that are directly obtained from specimens to study shape variations. Kerschbaumer et. al (2008) compared these two techniques in discriminating three populations of the cichlid fish,

*Tropheus moori*. GM was proved to be more flexible in data acquisition with body shape analysis conducted using the semi-landmark approach while TM is only restricted to distances and ratios of distances. Dudzik (2019) applied the discriminant function and canonical variate analysis using the statistical software, Fordisc 3.1 (FD3) to identify similarities in craniometric dimensions between Asian and Hispanic groups. GM was applied for a detailed analysis of the morphological similarities of both groups which revealed better identifications of overlapping cranium dimensions across populations.

The GM analysis can either be landmark-based or outline-based which uses the outline of a specimen (Richtsmeir et. al, 2002). Changbunjong et. al (2016) compared both approaches to distinguish three species of the *Stomoxys* adult flies based on their wing geometry. The wing shape was distinguished with significant differences in Mahalonobis distance, and the study concluded that both GM methods are useful in observing morphological distinctions of vectors among the three species of adult flies. Murphy and Garvin (2017) performed elliptical Fourier analysis on two-dimensional images of the left lateral, posterior and superior cranial views from 198 black and white individuals from the United States crania to assess population and sex variation. The aforementioned study observed that the outline analysis that incorporates multiple nonmetric traits into a single statistical analysis may result in more objective and accurate means of ancestry classification.

This study aims to incorporate functional data analysis (FDA) by representing the landmark coordinates used in GM analysis in the form of functions for shape analysis based on the landmark data obtained from the craniodental shape of three species of shrews. FDA refers to a branch of statistics that analyses and interprets data that exists on a continuum such as curves. Each sample elements are considered as a function under the FDA framework, which often defines time, spatial location or wavelength as the physical continuum. According to Musser, 2022, shrews are small insectivorous mammals with short limbs and tails that vary among species. Shrews have small eyes with moderately large, rounded ears (Musser, 2022). This study focuses on the variation among three shrew species: *Crocidura malayana* (*C. malayana*), *Crocidura monticola* (*C. monticola*) and *Suncus murinus* (*S. murinus*). *C. monticola,* also known as the Malayan shrew, belongs to family Soricidae. This small shrew species can be found in subtropical or tropical dry forests in Malaysia and Thailand. *C. monticola* is another small shrew species (weighs less than 8 grams) which belongs to the family Soricidae. This species can be found in primary and secondary montane forests in Malaysia and Indonesia (Peters, 1870). *S. murinus* (Asian house shrew) is originated from the Indian subcontinent and is commonly found in household communities and forest environments. It is one of the largest members of the genus *Suncus* of family Soricidae.

In this work, FDA is employed to analyze the image and shape data in the form of functions. Functional and shape analysis require tools to perform statistical analysis on signals, curves, or even more complex objects while being invariant to certain shape-preserving transformations (Guo et.al, 2022). To ensure that the functions are well-aligned for geometric features (peaks and valleys), curve registration (Ramsay and Li, 1998; Srivastava et. al, 2011) or functional alignment (Ramsay, 2006) are applied to warp the temporal domain of functions (Guo et. al, 2022). Epifanio and Ventura-Campos (2011) proved that the use of the FDA framework supersedes other approaches such as the landmark-based approach or even the set theory approach on principal component analysis (PCA) using a well-known database of bone outlines. FDA based on the landmark method aligns special features in functions or derivatives to their average location and then smooth to the location of the feature (Kneip and Gasser, 1992; Kneip and Gasser, 1995).

The landmarks obtained from the craniodental shapes of three species of shrews are represented in the form of functional data. This data is used to perform multivariate functional principal component analysis (MFPCA) to observe variation among the three shrew species and compared with the classical PCA. The principal component scores obtained from MFPCA (MFPC scores) were then reconstructed based on a truncated multivariate Karhunen-Loeve representation to produce predicted functions. These predicted functions were then applied to the classification models such as the naïve Bayes (NB), support vector machine (SVM), random forest (RF) and generalized linear model (GLM) models to examine the performance of classification accuracy of the three species of shrews. The NB classifier is based on Bayes' theorem and assumes that the attributes in a dataset are conditionally independent, given its class (Webb, 2011). SVM breaks down the multi-class classification into multiple binary classification cases by using a hyperplane that will separate data into their potential categories. RF consists of decision trees that uses the bootstrapping method to overcome overfitting on the training data to achieve better predictive accuracy (Denisko and Hoffman, 2018). GLM distinguishes and estimates responding variables following exponential distributions. These classifiers are commonly used in many studies related to species identification and classification tasks (refer to Bellin et.al, 2021; Macleod, 2018; Hernández-Ramírez and Aké-Castillo, 2014).

This paper is organized as follows: Section 2 gives a description on the shrew landmark data. Section 3 gives a detailed explanation on the methodology used based on the FDA approach and its application in the shrew landmark data. The three classification models applied are also provided. Section 4 describes the comparison between the FDA approach and landmark-based approach, followed by discussion on the produced results by the three classification models. Section 5 concludes.

## METHODOLOGY

*Proposed framework*

The implementation of the three shrew species is divided into four main steps: Shrew skull image acquisition, landmark data acquisition, implementation of functional data analysis (FDA) and output of the classification models.

*Shrew skull image acquisition*

The skulls of *C. malayana, C. monticola and S. murinus* can be viewed from different angles, i.e., dorsal, jaws, ventral and lateral (Figure 1), depending on the shape of the specimen. However, the ventral view was excluded in this study because both ventral and dorsal have identical shapes (Abu et. al, 2018). A total of 90 specimens of three different shrew species *(C. malayana, C. monticola, S. murinus)* were retrieved from the Museum of Zoology, Universiti Malaya (UM), Kuala Lumpur, Malaysia. All the skulls extracted from each specimen were separately placed in small bottles for geometric morphometrics. The technique of capturing the digital images of skulls was based on the method described in Abu et al, 2018. Skull digital images were captured using Nikon D90 with 15x magnification and stored in the Tagged Image File Format (tiff) format with a resolution of 4288 x 2848 pixels. Adobe Photoshop CS6 was also used to improve the image quality.

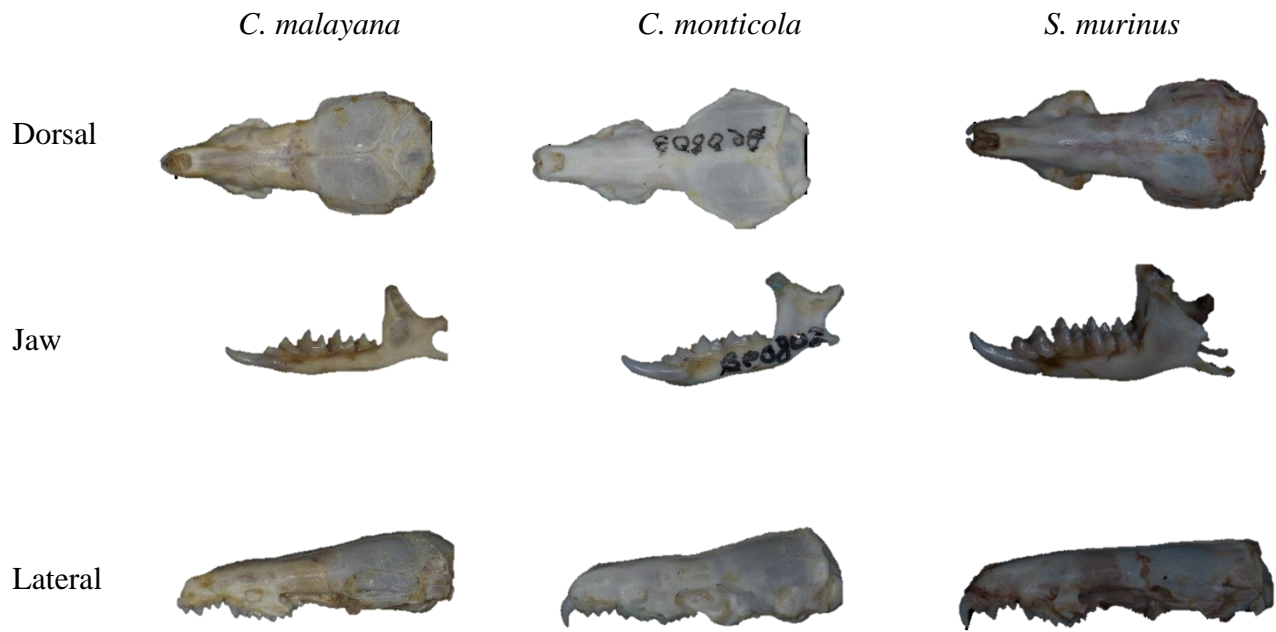

Figure 1 Digital skull images of dorsal, jaw and ventral views of *C. malayana, C. monticola and S. murinus*

*Landmark data acquisition*

After acquiring the images, TPSUtil32 is used to obtain the TPS files for all three views which will be used in TPSDig2 (Rohlf, 2013) for landmarking. Based on Figure 2, each view has different numbers of landmarks and semi-landmarks, i.e., dorsal (25 landmarks), jaw (50 landmarks) and lateral (47 landmarks). The statistical analysis of three views was performed in R version 4.2.1.

To use the geometric morphometric data, the raw coordinates obtained from the landmarks of all three craniodental views were processed using generalized Procrustes analysis (GPA) for optimal registration using translation, rotation, and scaling using the *gpagen* function in the *geomorph* package (Adams, et al, 2013). According to McCane, 2011, outline methods produce useful and valid results when suitably constrained by landmarks. This leads to the main idea of this work to incorporate the FDA approach to observe the separation among the three shrew species.

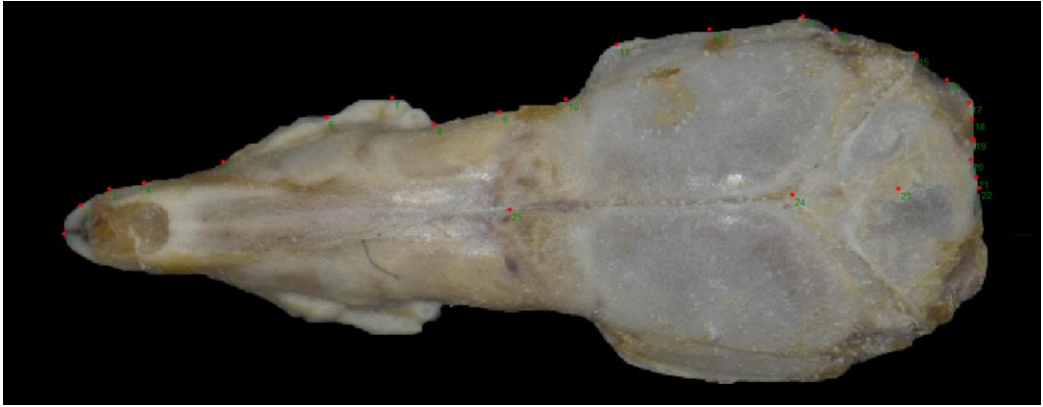

(a)

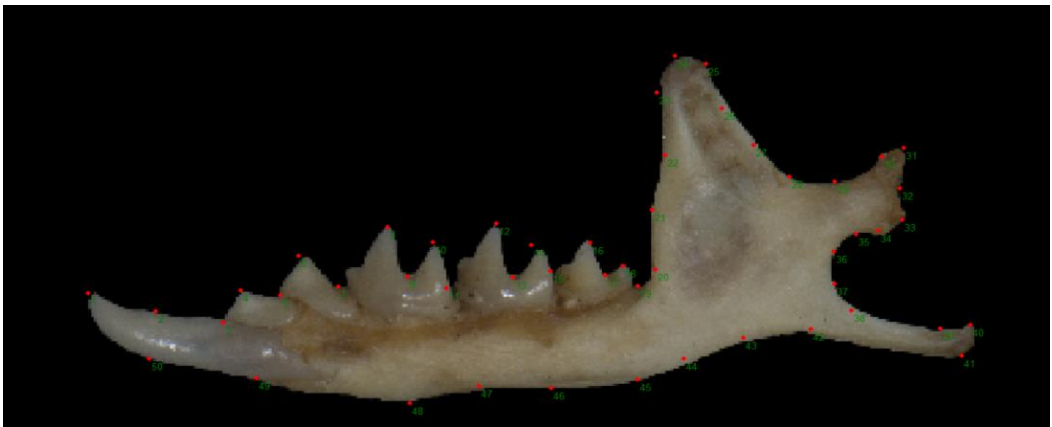

(b)

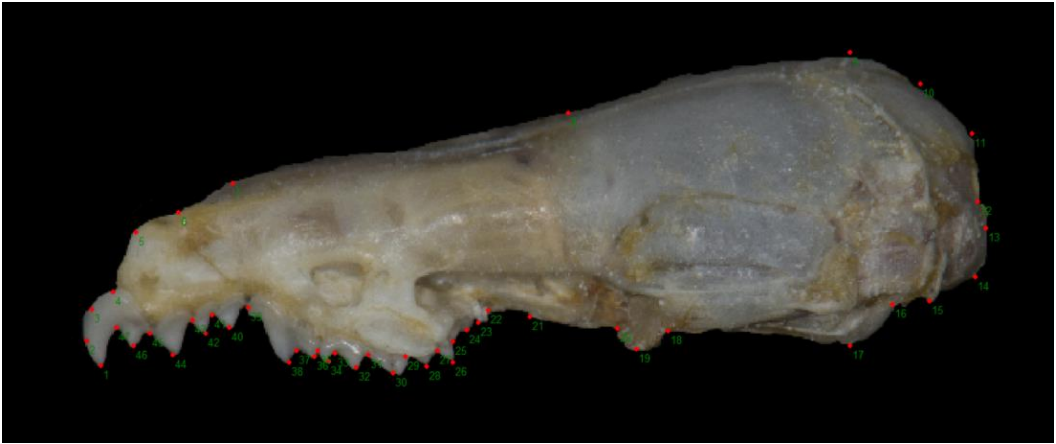

(c)

Figure 2 Digital images of the shrew skull from three views: (a) dorsal (b) jaws (c) lateral

*Implementation of functional data analysis (FDA)*

FDA is an approach used to analyze raw data with some dynamic in time, space, or more complex data as functions. In this study, the standardized coordinates from GPA were used to consider the outline of the shapes for these three views. Let $\{(x_i(t_1), y_i(t_1)), \ldots, (x_i(t_\tau), y_i(t_\tau))\}$ , $i = 1, \ldots, n$ be the collected data (standardized coordinates for one craniodental view) for $n$ specimens and $\tau$ is a p-tuple of $d_1, \ldots, d_p$ dimensional vectors and not scalar.. These discrete curve observations were converted into continuous functions using the *funData* package (Happ-Kurz, 2018) in R. Based on Happ-Kurz,2021, the functional data is sampled on a fine grid, $T \subset \mathbb{T}$, where $\mathbb{T}$ is the 1-dimensional domain, producing the sampling points in the domain as a numeric vector. The functional data $(X_i(t))_{i=1,\ldots,n}$ represents the landmark points as a functional data with the $n = 90$ specimens as observations. The functional data for each angle is combined as a multivariate functional data, with 90 observations defined on 2-dimensional domains. The 90 outlines each yields a vector of 50 coordinates (25 landmarks in 2 dimensions) for the dorsal view, a vector of 100 coordinates (50 landmarks in 2 dimensions) for the jaw view, and a vector of 94 coordinates (47 landmarks in 2 dimensions) for the lateral view.

The MFPCA estimates were computed using the *MFPCA* package on the functional data, based on their univariate counterparts (Happ-Kurz, 2018). This function calculates MFPCA based on observations that are independently and identically distributed (multivariate functional data obtained from the landmarks). MFPCA is a dimension reduction tool to transform sampled curves to represent the patterns of the variability of the curves. This is a more natural way to represent a multivariate functional data as they share the same structure as each observation (Happ and Grevens, 2015). The most common basis expansion on 1-dimensional domain, *uFPCA* (univariate FPCA) (Happ-Kurz, 2020) was applied on the 90 observations based on the PACE (principal components analysis through conditional expectation) approach (Yao et. al, 2009). Penalized splines are applied to smooth covariance surface (Di et. al, 2009) using the *refund* package (Happ-

Kurz, 2020). In MFPCA, vectors which are no longer considered as PCs are replaced by functions. Let $\{x_1(t), \ldots, x_j(t)\}$ be the set of pairs of observed multivariate functions. The average functions are obtained by centering. The mean function ($\bar{X}(t)$ is defined as:

$$\bar{x}(t) = \frac{\sum_{j=1}^{n} x_j(t)}{n}.$$

The centered functional data is obtained by

$$\left(X^{*(j)} = x_j(t) - \bar{x}(t)\right)_{j=1,\ldots,n}.$$

For each element $X^{*(j)}$, a univariate FPCA is estimated based on the observations $x_1^{(j)}, \ldots, x_n^{(j)}$. This results in the estimated eigenfunctions, $\widehat{\emptyset}_m^{(j)}$ and scores $\hat{\xi}_{i,m}^{(j)}$, $i = 1, \ldots, n, m = 1, \ldots, m_j$ for suitably chosen truncation lags $m_j$. The matrix $\xi \in \mathcal{R}^{n \times m_+}$, where m+ is the positive eigenvalues of the block matrix $Z$ and each row $(\hat{\xi}_{i,1}^{(1)}, \ldots, \hat{\xi}_{i,m_1}^{(1)}, \ldots, \hat{\xi}_{i,1}^{(p)}, \ldots, \hat{\xi}_{i,m_p}^{(p)})$ contains all estimated scores for a single observation. An estimate $\widehat{Z} \in \mathcal{R}^{m_+ \times m_+}$ of the block matrix $Z$ is given by $\widehat{Z} = (n-1)^{-1} \xi^T \xi$. A matrix eigen analysis for $\widehat{Z}$ is performed, resulting in eigenvalues $\hat{v}_m$ and orthonormal eigenvalues $\hat{c}_m$. The estimates for the eigenfunctions are given by:

$$\hat{\psi}_M^{(j)}(t_j) = \sum_{n=1}^{m_j} [\hat{c}_m]_n^{(j)} \widehat{\emptyset}_n^{(j)}(t_j), t_j \in \tau, M = 1, \ldots, m^+,$$

and multivariate scores are calculated using:

$$\hat{p}_{i,M} = \sum_{j=1}^{p} \sum_{n=1}^{m_j} [\hat{c}_m]_n^j \hat{\xi}_{i,n}^{(j)} = \xi_{i,\cdot} \hat{c}_m.$$

Based on Happ-Kurz, 2021, this function provides the MFPCA, $\hat{\psi}_M^{(j)}(t_j)$, associated eigenvalues $\hat{v}_1 \geq \cdots \geq \hat{v}_m > 0$ and the individual scores $\hat{p}_{iM} = \langle x_i, \hat{\psi}_M \rangle$. The results obtained were later reconstructed based on a truncated multivariate Karhunen-Loeve representation to produce predicted functions. The functions provide information on the estimated MFPCA scores given the weight of each observation's unit $t$ for its corresponding estimated multivariate eigenfunction, $\hat{\psi}_M$ (Lam and Wang, 2022). A detailed explanation of all propositions is available in Happ and Greven, 2016. The predicted MFPC scores from the landmarks were then applied into the classification models to distinguish among the three species of shrews and the results were compared with the classical PCA. The principal component scores were used for both approaches as they are capable to achieve low classification errors despite a major reduction of data (Howley et. al, 2005).

*Classification Models*

Machine learning has been extensively used in morphometric studies for classification and identification tasks (Tan, et al., 2018). NB, SVM, RF and GLM were chosen as classification models as these models were commonly used in many classifications related studies. Van der Plat et. al, 2021 applied NB and RF classifiers for species classification in plant genetic resources collections. NB and SVM are chosen as classifiers to classify species with DNA Barcode sequence (Weitschek et.al, 2014). GLM was one of the chosen classifiers to observe species distribution data at three fine scales: fine (Catalonia), intermediate (Portugal) and coarse (Europe) (Thuiller et. al, 2003). The performances of the NB, SVM, RF and GLM methods on classification of species among the shrews were assessed using the principal component scores from functional data (MFPCA) and classical PCA scores. This was done using the *e1071, MASS and caret* packages in R. The combined analysis of all three views and each separate view was performed. Monte Carlo simulation was performed with 20 iterations to observe the possible output of each model. A brief description of these classification models is provided as follows:

*The Naïve Bayes (NB) Classifier*

The NB classification model is a classifier used to estimate the posterior probability, to provide a mechanism that utilizes predictors of the training data (Webb, 2011). NB classifiers are trained to use the first two predicted principal component scores as predictor variables and the three species of shrews as class labels. This is done for the combination of all three views and separately and their performance measures are tabulated and compared. The process is performed using the predicted MFPCA scores.

*The Support Vector Machine (SVM) Classifier*

SVM addresses a multi-class problem as a single "all-together" optimization. SVM is used to find a hyperplane in the 2-dimensional space that will separate the scores to their potential species. For multiclass classification, the problem is split into multiple binary classification cases. In this approach, each classifier separates the points of two different species and combines all one-vs-one classifiers which leads to a multiclass classifier.

The Random Forest (RF) Classifier

RF uses decision trees on all the predictor variables of the training data (MFPCA scores) using the three species of the shrews as classification category to improve the predictive accuracy. The model is then assessed using the test data and the results of the performance measures are tabulated. Generalized Linear Model (GLM)

GLM estimates the response variables following exponential distributions. The three species of shrews were used in the response columns and the predicted principal components scores act as the main effects of the predictor variables.

## RESULTS

MFPCA using the functional data of all views combined indicated a total of 31 MFPCs. The first two MFPCs explained 81.56% of the total variation in the species of shrews. A distinct cluster among the species of shrews when using the predicted MFPC scores (Figure 3) is also observed. The classical PCA yields 89 principal components, where the first two PCs explained 62.94%. Although *S. murinus* does seem to be well separated in the classical approach, the method could not clearly distinguish the other two shrew species (Figure 4).

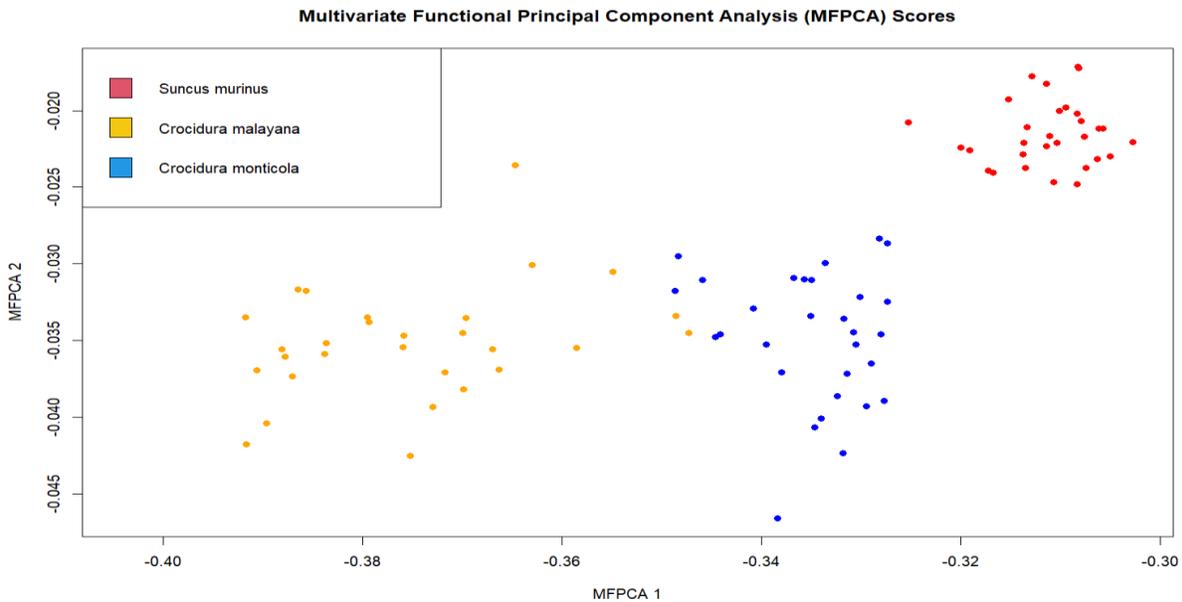

Figure 3 MFPCA plot for all three views ( dorsal, jaw and lateral combined)

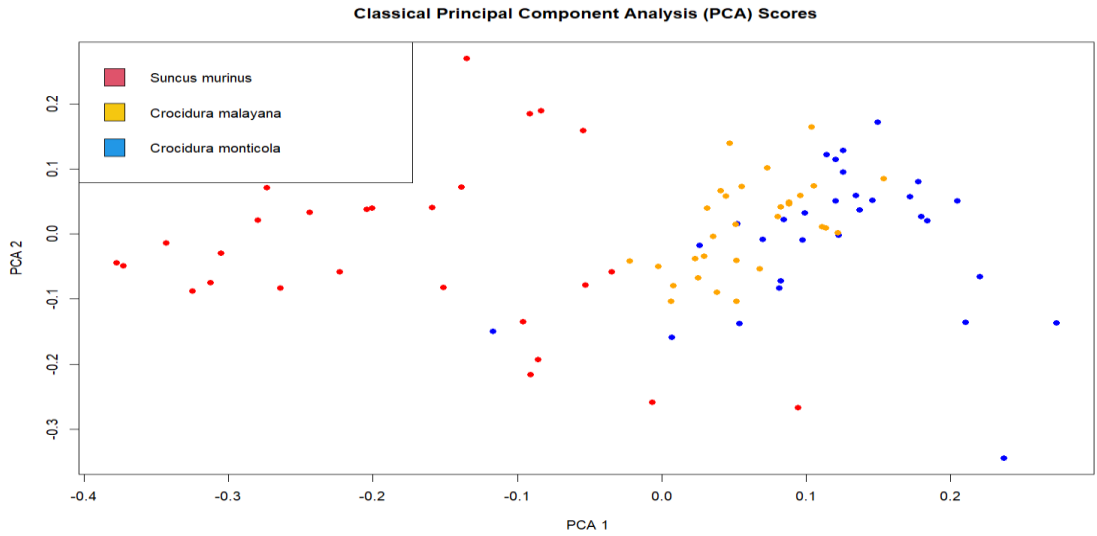

Figure 4 Classical PCA plot for all three views ( dorsal, jaw and lateral combined)

Thus, employing the FDA approach with functional principal component analysis has potential in examining the species variation of the shrews. When PCA is separately conducted on each view based on the FDA and classical approaches, the dorsal view (Figure 5) gives the best separation for the three shrew species compared to the other two views (Figure 6 and Figure 7) for both approaches.

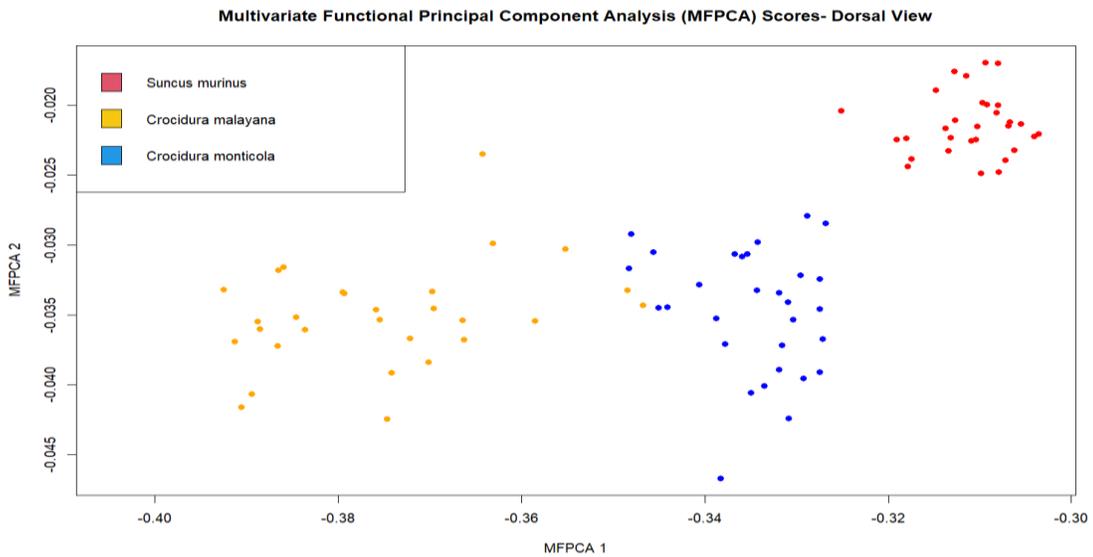

(a)

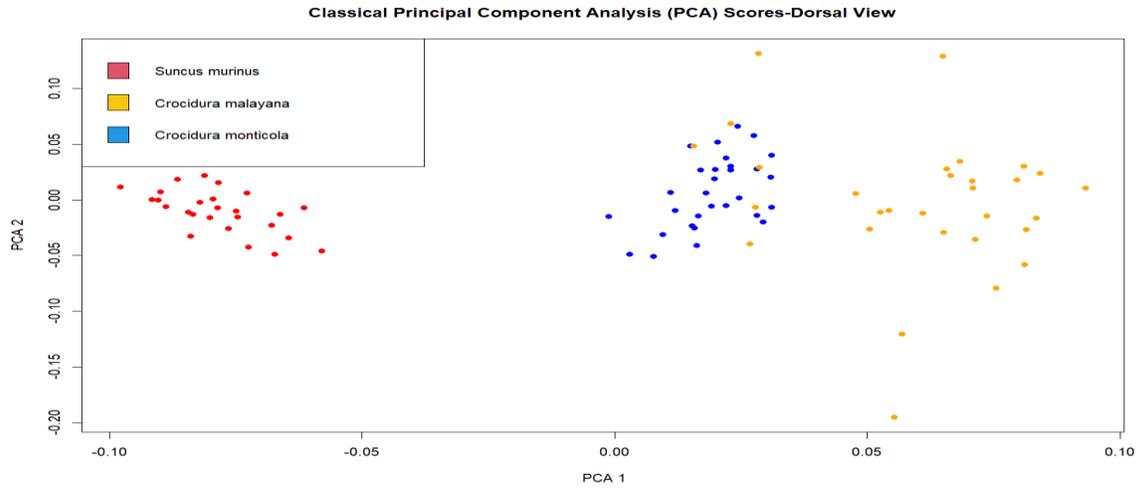

(b)

Figure 5 (a) MFPCA plot and (b) classical PCA plot for dorsal view

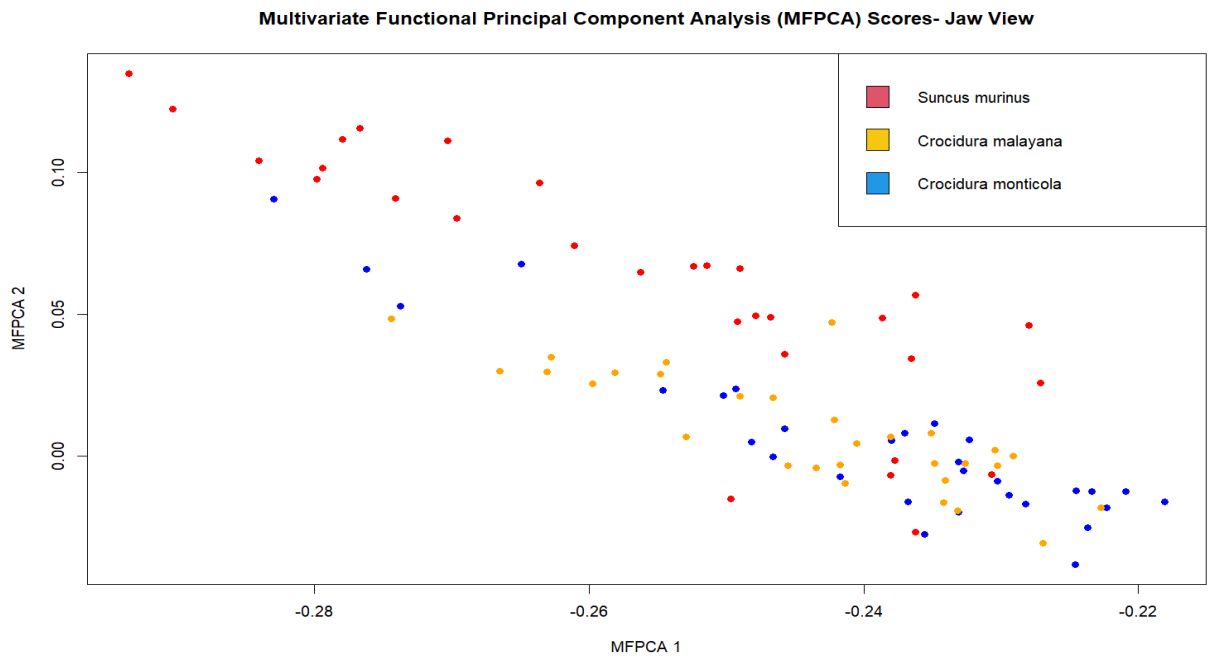

(a)

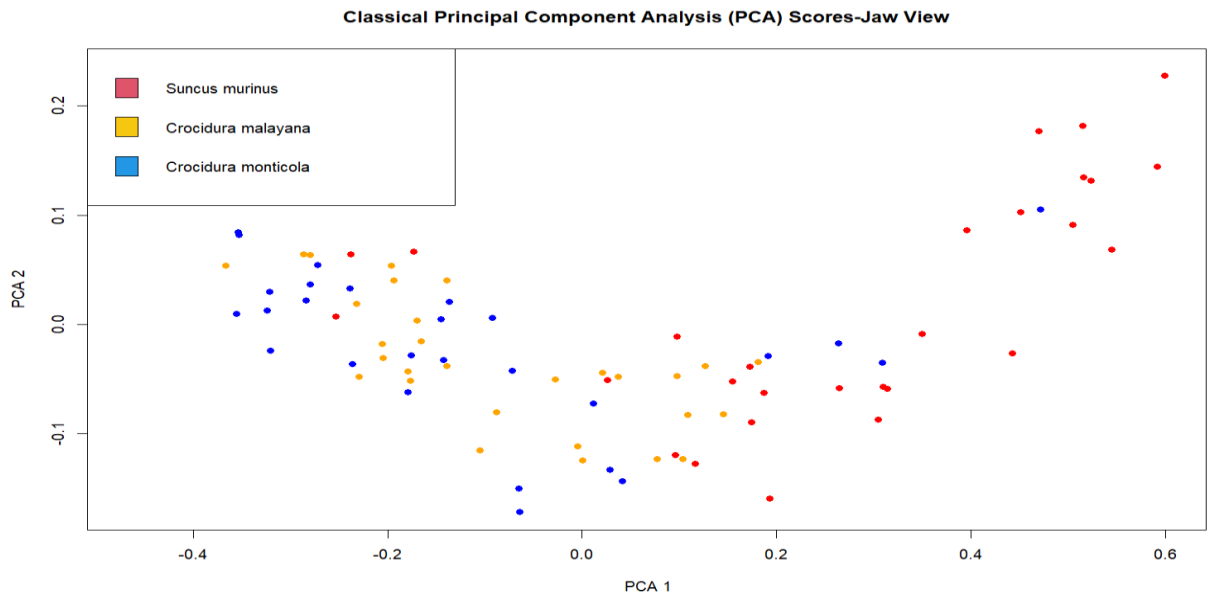

(b)

Figure 6 (a) MFPCA plot and (b) classical PCA plot for jaw view

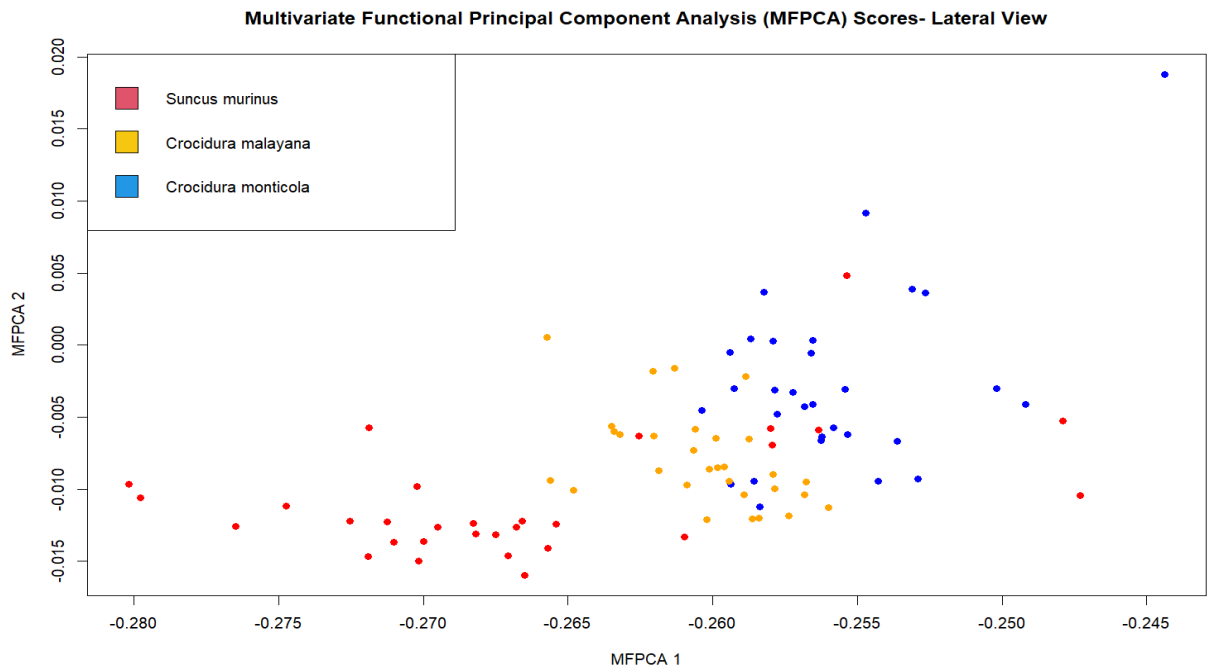

(a)

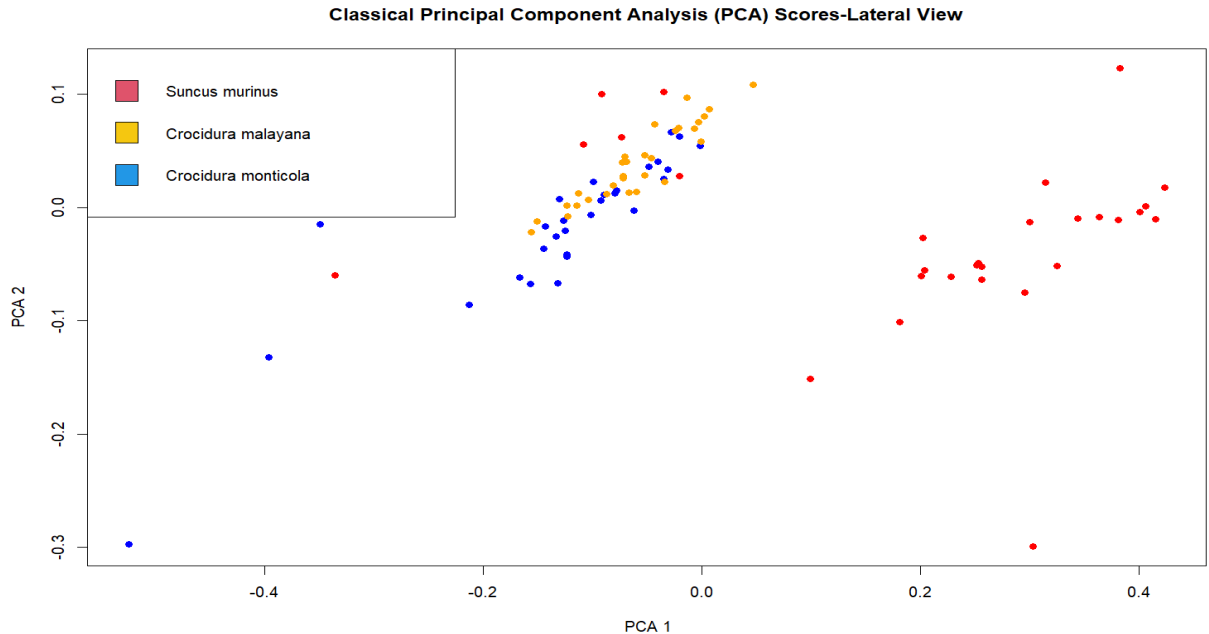

(b)

Figure 7 (a) MFPCA plot and (b) classical PCA plot for lateral view

The dorsal view yielded a total of 10 fPCs and the first two MFPCs explained 86.4% of the variation among the species. The classical approach yields 46 PCs and the first two explained 59.24% of variation. The predicted MFPCA results gave a better separation among the three shrew species compared to the classical method.

There are 11 MFPCs for the jaw view where the first two MFPCs explained 89.31% of the variation in the species. There is a total of 89 classical PCs for the jaw view where the first two components explained 73.13% of the variation. As for the lateral view, there is a total of 10 MFPCs and the total variation in species explained by the first two MFPCs is 90.9%. Out of 89 PCs, the first 2 PCs of the classical approach for lateral view explained 74.29% of total variation. Though *S. murinus* is somewhat separated (Figure 6(b) and Figure 7(b)), the jaw view and lateral view give poor classification of all three species for the classical approach. A slight improvement of species separation can be observed in the FDA approach (Figure 6(a) and Figure 7(a)) for both views. The performance of the classification models based on individual craniodental views and the combination of all three is evaluated using the first two PC scores of both the FDA and classical approaches as the first two principal components of all the craniodental views lie within the general rule of thumb threshold of 80% in the FDA approach.

Table 1 shows an overall improvement in results for all the classification models when the FDA approach is applied compared to the classical method.

Table 1: Evaluation of model performance in MFPCA scores and PCA scores for all three views (dorsal, jaw and lateral combined)

| Classifiers | MFPCA | | Classical PCA | |
| --- | --- | --- | --- | --- |
| | Mean | Standard Deviation | Mean | Standard Deviation |
| **NB** | 0.969 | 0.036 | 0.806 | 0.066 |
| **SVM** | 0.957 | 0.040 | 0.776 | 0.075 |
| **RF** | 0.948 | 0.052 | 0.780 | 0.079 |
| **GLM** | 0.968 | 0.029 | 0.806 | 0.064 |

The mean classification accuracy obtained from the NB and SVM models noticeably outperform the other two classification models for both approaches. Based on Table 2 and Table 3, all four classification models produce excellent results in terms of classification accuracy when only the principal component scores of the dorsal view are used for both classical and FDA approaches.

Table 2: Evaluation of model performance in MFPCA scores for all three views (separated)

| Classifiers | Mean | | | Standard Deviation | | |
| --- | --- | --- | --- | --- | --- | --- |
| | Dorsal | Jaw | Lateral | Dorsal | Jaw | Lateral |
| **NB** | 0.952 | 0.572 | 0.791 | 0.036 | 0.072 | 0.067 |
| **SVM** | 0.957 | 0.600 | 0.796 | 0.040 | 0.086 | 0.047 |
| **RF** | 0.948 | 0.530 | 0.772 | 0.052 | 0.076 | 0.084 |
| **GLM** | 0.954 | 0.586 | 0.797 | 0.037 | 0.093 | 0.072 |

Table 3: Evaluation of model performance in PCA scores for all three angles (separated)

| Classifiers | Mean | | | Standard Deviation | | |
| --- | --- | --- | --- | --- | --- | --- |
| | Dorsal | Jaw | Lateral | Dorsal | Jaw | Lateral |
| **NB** | 0.931 | 0.541 | 0.654 | 0.040 | 0.077 | 0.076 |
| **SVM** | 0.935 | 0.524 | 0.639 | 0.067 | 0.074 | 0.088 |
| **RF** | 0.898 | 0.591 | 0.652 | 0.048 | 0.093 | 0.072 |
| **GLM** | 0.933 | 0.557 | 0.665 | 0.040 | 0.107 | 0.090 |

GLM gives the best classification accuracy compared to the other three models for the dorsal view. The jaw view provides the least favorable classification accuracy compared to the other two craniodental views for both approaches. Improvement in classification accuracy for all models can be observed when the FDA approach is implemented. Overall, the dorsal view seems to show promising results to observe variation among the three shrew species compared to the other two craniodental views.

## DISCUSSION

A comparison of the FDA and classical approaches was done to study the classification of *S.murinus*, *C. monticola* and *C. malayana* using craniodental landmarks extracted from the skull images of shrews. The principal component scores which were projected onto orthonormal eigenfunctions capture prominent directions, thus improving classification (Lee, 2004). The findings of this study revealed the existence of distinct clusters of the shrew species when the standardized landmarks of the three craniodental views combined are converted into functional data, rather than being discretized in point sets. When the classical landmark-based approach is used, it can be difficult to standardize the selections, thus leading to drastically differing results (Srivastava and Klassen, 2016). In this case, the FDA approach can be a more natural solution as the boundaries of the objects are treated as continuous curves, thus better matches the features across curves (Srivastava and Klassen, 2016).

Performing PCA on functional data is advantageous of regularization which is an issue found in many datasets as the approach can better reveal if some type of smoothness is required on the principal components (Ramsay and Silverman, 2005). MFPCA reduces dimensionality by projecting the functional landmark data onto the set of orthonormal basis functions which induces the uniqueness of the MFPCA scores for each observation to improve classification accuracy. Predicted MFPCA scores based on a truncated multivariate Karhunen-Loeve representation reduces the dimension to the first few components (Muiller and Stadmuller, 2005).

As shown in Figure 5(b), PCA based on the standardized landmark data does not give a fully comprehensible presentation of the structure variability compared to the FDA approach. When the three craniodental views were individually examined (Figures 5, 6, and 7), the dorsal view showed the clearest separation among the three shrew species based on both approaches. The MFPCA and PCA results were also verified using the linear measurements of the three shrew species. The similarities of the measurements among the three species were observed. Based on the observation, the linear measurements of the dorsal view of these specimens do show that the GLS, BB and IOB vary greatly for the three species which supports the results obtained in Table 6. These linear skull measurements are based on Omar, 2013.

Table 6: Ranges of skull measurements (in millimeters) for the dorsal view of all three species.

| Dorsal | | | |
|---|---|---|---|
| Character | *S. murinus* | *C. malayana* | *C. monticola* |
| Greatest length of skull (GLS) | 29-35 | 22-25 | 16.2-18.1 |
| Braincase breadth (BB) | 12-14 | 8-10.5 | 7.1-8.3 |
| Interorbital breadth (IOB) | 5-6.2 | 4-5 | 3.5-4.2 |
| Jaws | | | |
| Mandibular Length (MAL) | 11-13.7 | 9.1-10.4 | 6.2-7.6 |
| Lower tooth row length excluding first incisor (IM3I) | 8.1-9.5 | 6.7-7.6 | 4.6-5.4 |
| Lateral | | | |
| Condyle to glenoid length (CTG) | 10.6-13.9 | 8.4-8.9 | 6.4-7.2 |
| Post-palatal depth (PPD) | 5.2-6.5 | 4-4.3 | 3.0-3.6 |
| Rostral length (ROL) | 11.7-14.1 | 8.4-10.2 | 6.0-7.2 |

The least favorable classifications are observed from the jaw view (Figure 6), although there is an implementation of the FDA approach shows some improvement. This may be due to the similarities in *C. monticola* and *C. malayana* as they belong to the same genus. The edges of the molar region tend to be similar for both species. The horseshoe effect present in the classical approach (Figure 6(b)) may indicate species turnover along environment gradients (Morton et.al, 2017). This effect has been commonly observed in ecological ordination obtained by the classical PCA (Podani and Miklos, 2002). The plots of the predicted MFPCA scores reveal the presence of functional manifolds where the horseshoe effect is noticed (Wang et al, 2015). The lateral view also indicates an overlap between the two species. This can be due to the similarity of the back curvature between the two as the region tends to be flat and a little sharp for *S. murinus*.

To ensure consistent results for the classification models, the same identification numbers were used for both training and testing data (for combination of all views and individual views) for all four classification models. Considering this study was based on functions of craniodental curve based on landmarks, there is a great improvement in classification rate for all four classification models when the FDA approach is applied. The dorsal view gives the best rate of classification accuracy among the three views.

## CONCLUSIONS

This study proposed the FDA approach on landmark data to represent the shapes of the of the dorsal, lateral and jaw of shrew skulls. The approach was applied to classify three shrew species (*S.murinus*, *C.monticola* and *C. malayana*) and was compared with the classical approach using four classifiers (NB, SVM, RF and GLM). The results confirmed that the FDA approach has improved the classification among the three species compared to the classical landmark-based approach. It was also revealed that the dorsal view may be the best representation in classifying the three species for both approaches. The FDA approach is advantageous as it treats the landmark coordinates as a curve rather than discrete points, thus providing more information contained in a spatial region. The adoption of functional forms in morphometrics has particularly strong applications in biological shape analysis. Hence, FDA-related morphometrics is a potential tool in enhancing morphometrics research. Future studies based on outline analysis using the FDA approach on the images of these craniodental views of shrews can also be considered to improve classification accuracy.

## ACKNOWLEDGEMENTS

This work was supported by the Universiti Malaya, Faculty Research Grant [GPF088-A2020].

.